\documentclass{osa-article}

\journal{oe}
\usepackage{lineno}
\usepackage{url}
\usepackage{hyperref}
\usepackage{placeins}
\usepackage{textcomp}
\usepackage{amsmath}
\usepackage{graphicx}
\usepackage{epsfig}
\usepackage{epstopdf}
\usepackage{pdfpages}
\begin{document}

\title{Phase-quadrature quantum imaging with undetected photons}

\author{Björn Erik Haase,\authormark{1,2,*} Joshua Hennig,\authormark{1,2} Mirco Kutas,\authormark{1,2} Erik Waller,\authormark{1} Julian Hering,\authormark{2,3} Georg von Freymann,\authormark{1,2,3} and Daniel Molter\authormark{1}}

\address{\authormark{1}Fraunhofer Institute for Industrial Mathematics ITWM, Fraunhofer-Platz 1, \\67663 Kaiserslautern, Germany\\
\authormark{2}Physics Department and Research Center OPTIMAS, Technische Universität Kaiserslautern, \\67663 Kaiserslautern, Germany\\
\authormark{3}Opti-Cal GmbH, Erwin-Schrödinger-Straße 56, 67663 Kaiserslautern, Germany}

\email{\authormark{*}bjoern.erik.haase@itwm.fraunhofer.de}

\begin{abstract}
Sensing with undetected photons allows access to spectral regions with simultaneous detection of photons of another region and is based on nonlinear interferometry. To obtain the full information of a sample, the corresponding interferogram has to be analyzed in terms of amplitude and phase, which has been realized so far by multiple measurements followed by phase variation. Here, we present a polarization-optics-based phase-quadrature implementation in a nonlinear interferometer for imaging with undetected photons in the infrared region. This allows us to obtain phase and visibility with a single image acquisition without the need of varying optical paths or phases, thus enabling the detection of dynamic processes. We demonstrate the usefullness of our method on a static phase mask opaque to the detected photons as well as on dynamic measurement tasks as the drying of an isopropanol film and the stretching of an adhesive tape. 
\end{abstract}

\section{Introduction}

The quantum-optical measurement principle using so-called undetected photons enables access to spectral ranges that are not directly addressable with the employed detectors. The core principle is building a nonlinear interferometer, in whose crystals correlated photons (called signal and idler) are generated by spontaneous parametric down-conversion (SPDC) that show interference in the signal beams based on indistinguishability, if the idler beams are overlapped\cite{Wang1991, Zou1991}. A sample introduced in the idler beam path influences the phase and the visibility of the signal's interference. Thus, those measurement approaches enable the measurement of sample properties in the idler spectral range by solely detecting photons in the signal wavelength range \cite{Kutas2022}. This principle has gained an enormous attention in the last decade and has already been demonstrated in various spectral ranges and applied to various tasks as imaging\cite{Lemos2014,Kviatkovsky2020,Vanselow2020}, sensing\cite{Paterova2019,Kutas2020}, and spectroscopy \cite{Kalashnikov2016, Lindner2022,Kutas2021}. 

As in traditional interferometry, the phase and visibility of an interferogram have to be measured by changing optical path differences. If a dynamic process has to be measured, the acquisition of the interferogram has to be conducted in a comparatively short time, challenging this measurement principle. Recently, the holography principle has been adapted to nonlinear interferometry, which allows to reduce the measurements to a set of known phases, but still requires the sequential acquisition of several measurements \cite{Topfer2022}. 

In this work, we introduce a phase-quadrature detection in a nonlinear interferometer using various polarization optics to simultaneously acquire four images at phases with a shift of $\pi /2$ to each other, inspired by similar concepts already implemented in traditional linear interferometers \cite{Hu2015, Wang2020}. By doing so, we obtain the phase and visibility image in a single image acquisition without the need of varying optical path lengths. We demonstrate measurements on a static phase mask opaque to the detected signal photons as well as dynamic processes as the drying of an isopropanol film and the continuous stretching of an adhesive tape. Most of the measurements are accompanied by classical measurement principles to verify the findings of the phase-quadrature measurement with undetected photons. Our concept is a further step towards the use of nonlinear interferometers for highly dynamic and time-critical measurements.

\section{Theoretical considerations on phase quadrature with undetected light}

Sensing with undetected idler photons relies on the indistinguishability of the path and origin of the detected signal photons. These signal and idler photons may differ and they can even have strongly non-degenerate frequencies \cite{Haase2019, Kutas2022}. The spatially varying signal photon detection probability $P_{\varphi}(x,y)$ depends on the losses $L(x,y)$ and the sample-induced phase $\phi (x,y)$ that is imprinted on the idler photons \cite{Topfer2022}. For the sake of simplicity, the spatial dependence is neglected in the following:

\begin{equation}
    P_{\varphi}\thicksim 1+\left(1-L\right)\cos\left(\phi +\varphi\right),
\end{equation}
with $\varphi$ being the global phase. From the analysis of $P_{\varphi}$, the spatial-introduced phase and losses of the photons can be evaluated.

Yet, to determine the complex transmission coefficient one needs to measure $P_{\varphi}$ for a set of various $\varphi$. A convenient possibility to measure them is by recording images with $\varphi = 0$, $\pi /2$, $\pi$, $3\pi /2$ and process these recordings. With these phases, $P_{\varphi}$ is
\begin{equation}
    \begin{split}
        P_{0} \thicksim1+(1-L)\cos\left(\phi\right) \hspace{24pt} P_{\pi /2} \thicksim1+(L-1)\sin\left(\phi\right)\\
        P_{\pi} \thicksim1+(L-1)\cos\left(\phi\right) \hspace{20pt} P_{3\pi /2} \thicksim1+(1-L)\sin\left(\phi\right)
    \end{split}\hspace{6pt}.
\end{equation}

With these four phase-dependent measurements, the phase introduced in the idler beam path can be calculated by
\begin{equation}
        \phi = \arctan\left({\frac{P_{3\pi /2}-P_{\pi /2}}{P_{0}-P_{\pi}}}\right).
        \label{eq:Phase}
\end{equation}

The interference visibility $V$ as the counterpart to the summed losses can be determined by
\begin{equation}
    V = 1-L = 2\cdot\frac{\sqrt{\left(P_{3\pi /2}-P_{\pi /2}\right)^{2}+\left(P_{\pi}-P_{0}\right)^{2}}}{P_{0}+P_{\pi /2}+P_{\pi}+P_{3\pi /2}}.
    \label{eq:Visibility}
\end{equation}

In this work, these four recordings are measured within a single image acquisition with the help of a series of phase plates.

\section{Experimental setup}

\begin{figure}[htb]
    \centering
    \includegraphics[width=\linewidth]{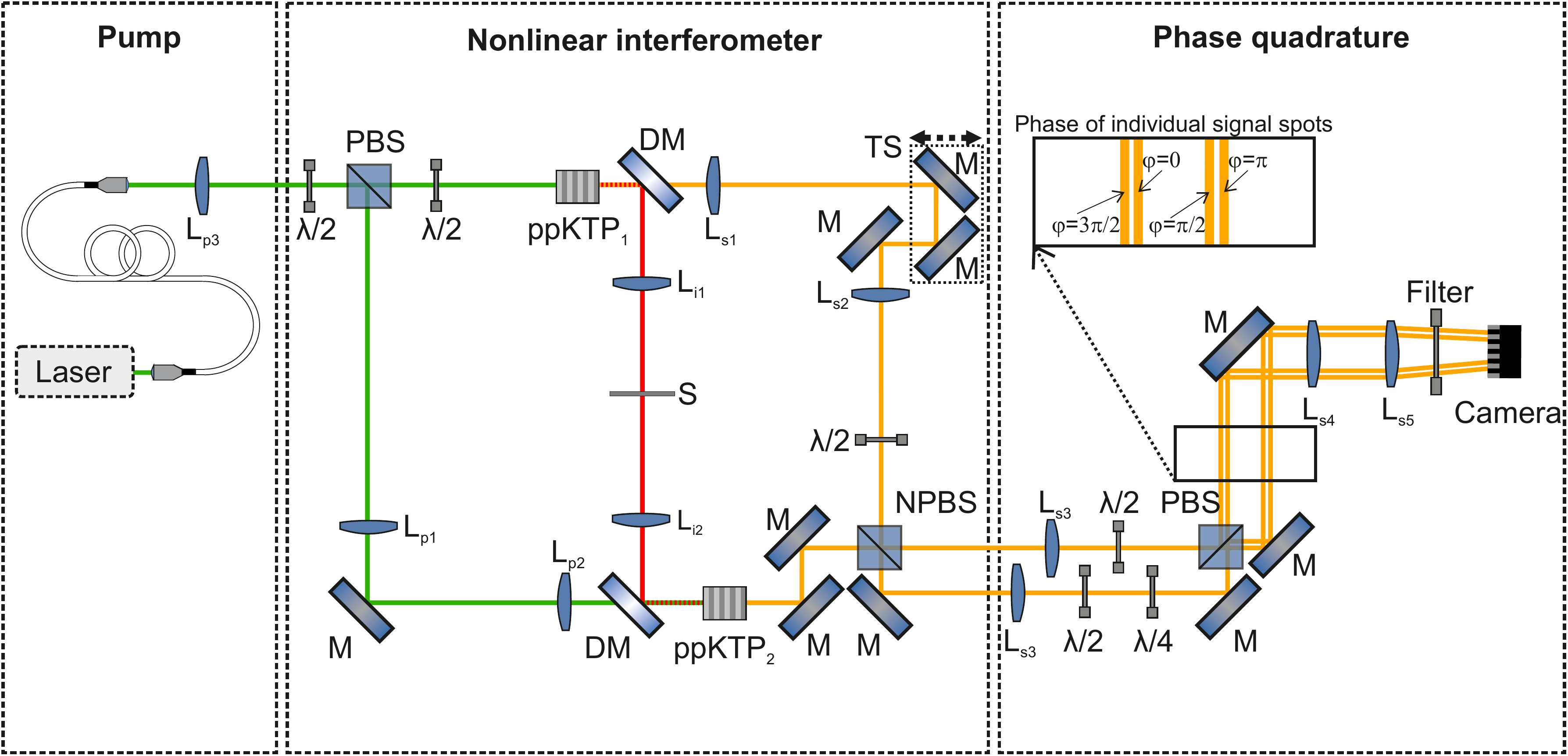}
    \caption{Experimental setup. Additional to the setup in \cite{Lemos2014}, a half-wave plate ($\lambda/2$) in front of the non-polarizing 50:50 beam splitter (NPBS) is added. Furthermore, the four quadratures are set by various wave plates ($\lambda$/2 and $\lambda$/4) and a polarizing beam splitter (PBS) behind them. L$_{\text{p}n, \text{s}n, \text{i}n}$: lens for pump (colored in green), signal (colored in yellow) and idler (colored in red) beams, ppKTP: nonlinear crystal, DM: dichroic beam splitter, S: sample, M: mirror, TS: translation stage. Lengths of optical paths are not to scale.}
    \label{fig:Setup}
\end{figure}

Our experimental setup shown in Fig.~\ref{fig:Setup} is based on a nonlinear Mach-Zehnder interferometer similar to the setup of Lemos \textit{et al.} \cite{Lemos2014}.

Starting with a pump laser at a wavelength of 532\,nm (green colored), we split its output with a polarizing beam splitter (PBS) and a preceding half-wave plate ($\lambda$/2). An additional half-wave plate rotates the polarization of the transmitted pump part by $\pi$/2 ensuring the same polarization in both interferometer arms. In both branches, one nonlinear periodically-poled potassium titanyl phosphate (ppKTP) crystal is placed. Triggered by SPDC, in the crystals some of the pump photons decay into an infrared idler photon with a wavelength around 1557\,nm (red colored) and a photon with the signal wavelength of about 808\,nm (yellow colored). The idler output of the first crystal is separated from the correlated signal radiation by means of a dichroic beam splitter (DM) and directed to the sample. Behind the sample, another DM directs the idler radiation into the second ppKTP crystal. The signal output of the second crystal is superimposed on the signal photons generated in the first crystal at a non-polarizing (50:50) beam splitter (NPBS). Various lenses focus the pump beam into the crystals and ensure that the beam waists are the same in both crystals and at the input ports of the NPBS, respectively. 

Compared to the setup of \cite{Lemos2014} there are two main differences. A piezo-controlled translation stage in the upper branch of the interferometer can adjust the delay between the two signal beams very precisely. It is solely used to ensure temporal overlap of the interfering signal beams and for calibration purposes, but not for the pursued phase-quadrature measurement technique. Additionally, a half-wave plate rotates the polarization in the first signal beam path by $\pi/2$, leading to a cross-polarization state. After spatially and temporarily superimposing the two signal beams, they do not yet interfere at the NPBS as they are distinguishable by their polarization. Both of the two cross-polarized outputs of the NPBS are rotated by a half-wave plate in a way that their polarization states are rotated by $\pi/4$ compared to the axis of the PBS. Since each signal beam contains contributions of both polarization states the PBS projects onto, the outputs of the PBS show the two signal beams' interference having a phase shift of $\pi$. To observe four different phases, the second output of the NPBS is additionally sent through a quarter-wave plate ($\lambda$/4) adding a phase shift of $\pi$/2 to one of the signal beams in this arm. Overall, this provides four outputs showing interference with global phases $\varphi =0$, $\pi$/2, $\pi$ and 3$\pi$/2.

To measure the entire signal at once with the same camera, the input ports of the PBS are transversely offset from each other, and the four output ports pass through a telescope with multiple mirrors between them to reduce their spot sizes before being sent to a single camera chip. To block the remaining pump radiation, several long-pass filters and a narrowband wavelength filter for the signal wavelength are placed directly in front of the camera. For the phase-quadrature measurement the samples are placed in the image plane between the two idler lenses L$_{\text{i1}}$ and L$_{\text{i2}}$ (here, the field of view (FOV) is about 11\,mm). Thus, depending on the optical thickness of the sample, an additional phase $\phi (x,y)$ is imprinted on the phase of the signal interference.

For all subsequent measurements, the nonlinear interferometer is pumped at a total pump power of about 60\,mW and the illumination time of the camera is 500\,ms. Phase stability of our setup can be maintained for several minutes.

\section{Results and discussion}
\subsection{Calibration}
\begin{figure}[h]
    \centering
    \includegraphics[width=0.8\linewidth]{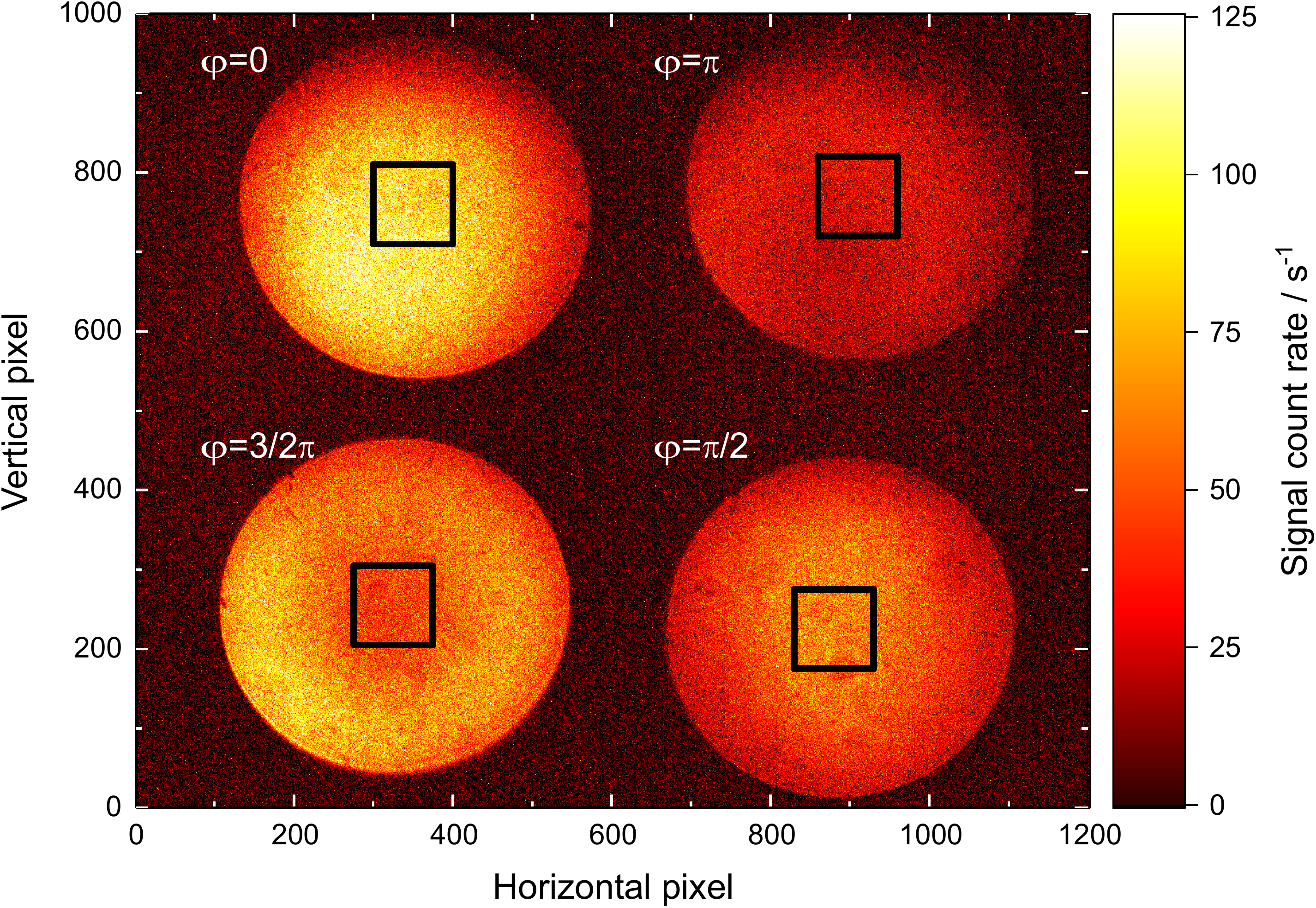}
    \caption{Signal spots of each of the four phase-quadrature branches. This raw image was recorded with a total pump power of 60\,mW and an integration times of 500\,ms at a delay of 0~µm with respect to Fig.~\ref{fig3_reference_waveform_phase}. The mean count rates within the squares are used to determine the interferograms and phase shown in Fig.~\ref{fig3_reference_waveform_phase}.}
    \label{fig2_calib_image}
\end{figure}

To align the setup and calibrate the four spots on the camera, interferograms of each of the four spots are recorded by acquiring images with varying delay. For this measurement, no object is placed in the beam path of the idler. 
One typical camera image of this calibration series is depicted in Fig.~\ref{fig2_calib_image}.

The interferograms are obtained by averaging the count rate within the marked area of 100 by 100 pixels in the center of the four spots. However, due to unequal performance of the employed optics in the quadrature beam paths, the separately calculated visibility values of the interferograms and the average count rate of the spots do not match. To compensate for this discrepancy, the individual pixel-wise count rates were corrected accordingly. Using eq.~\ref{eq:Phase}, the phase of the interference pattern can be calculated for each frame shot individually using the measured quadrature count rates. 

\begin{figure}[h]
    \centering
    \includegraphics[width=0.495\linewidth]{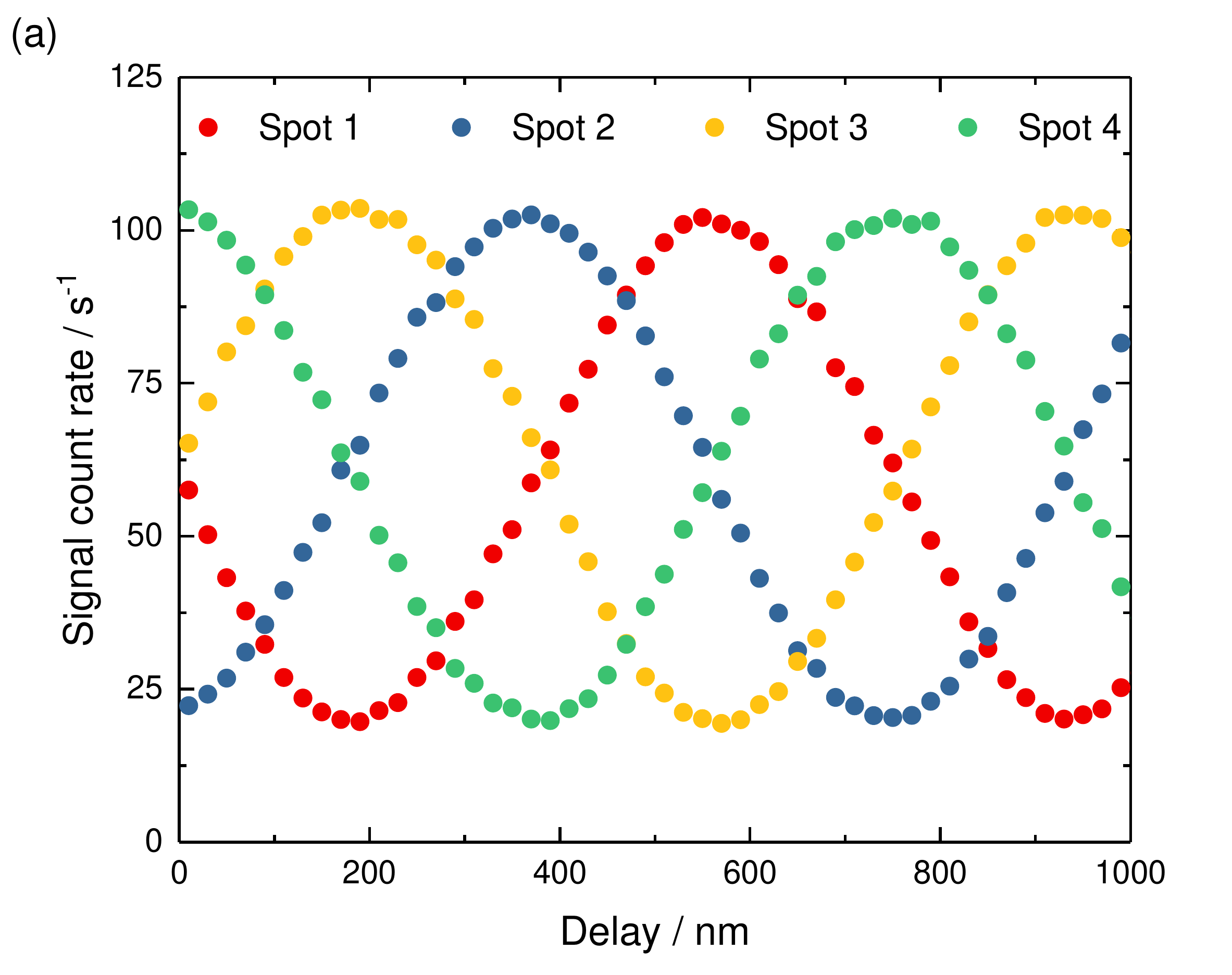}
    \hfill
    \includegraphics[width=0.495\linewidth]{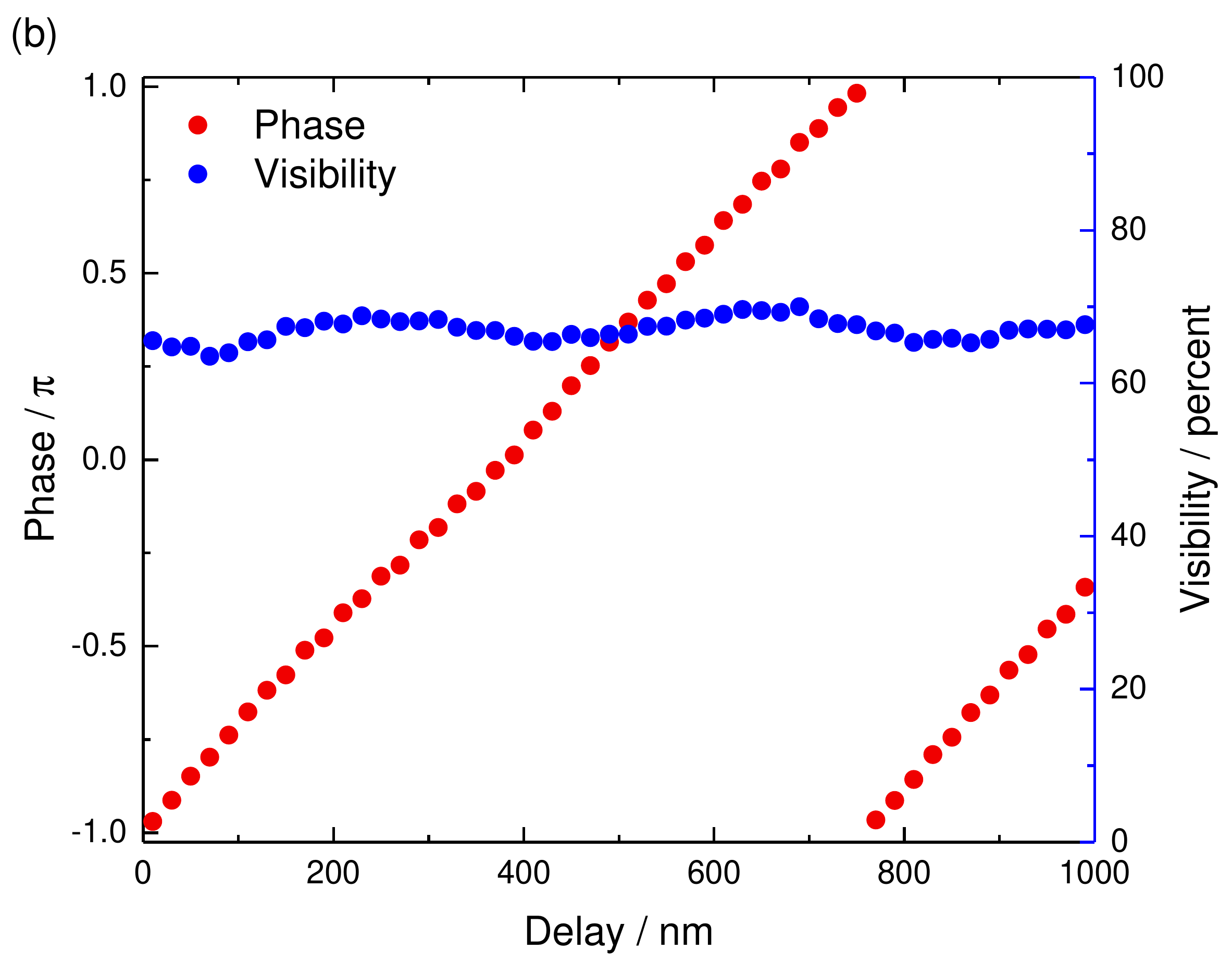}
    \caption{(a) Interferogram of each of the four signal spots and (b) phase and visibility as a function of a changed delay. The count rate of each spot is calibrated so that the mean count rate and the visibility of the interference is the same for all four spots.}
    \label{fig3_reference_waveform_phase}
\end{figure}

Along these lines the data shown in Fig.~\ref{fig3_reference_waveform_phase} are evaluated: The corresponding corrected interferograms of the four signal spots are depicted in Fig.~\ref{fig3_reference_waveform_phase}(a), the phase and visibility when moving the translation stage is plotted in Fig.~\ref{fig3_reference_waveform_phase}(b).

The individual phase obtained from single images shows a good correlation to the induced delay (coefficient of determination $r^{2}=0.9997$), while the visibility is reasonable constant (absolute standard deviation of 1.5\% at a mean value of 67\%). 

\subsection{Static phase mask}

\begin{figure}[h]
    \centering
    \includegraphics[width=0.495\linewidth]{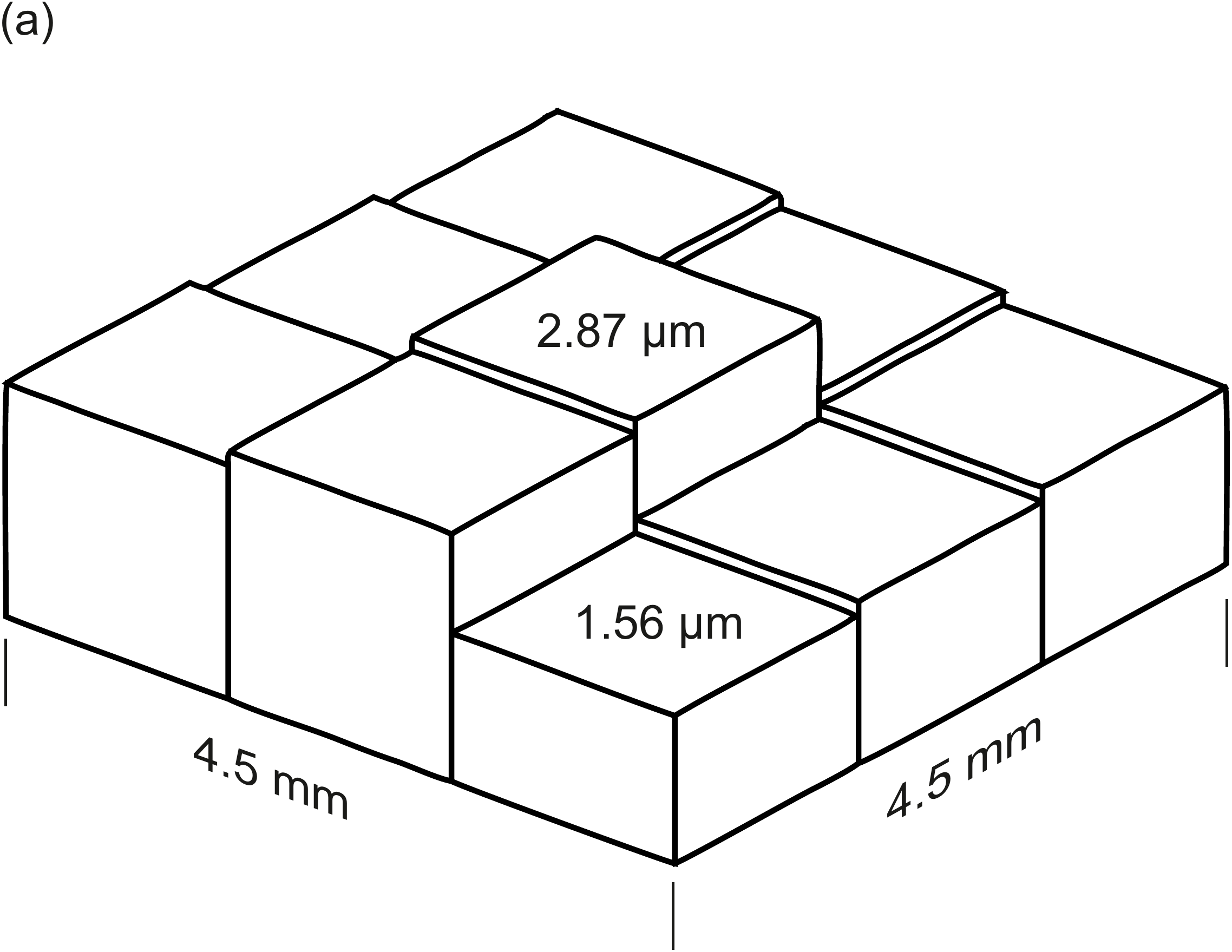}
    \hfill
    \includegraphics[width=0.495\linewidth]{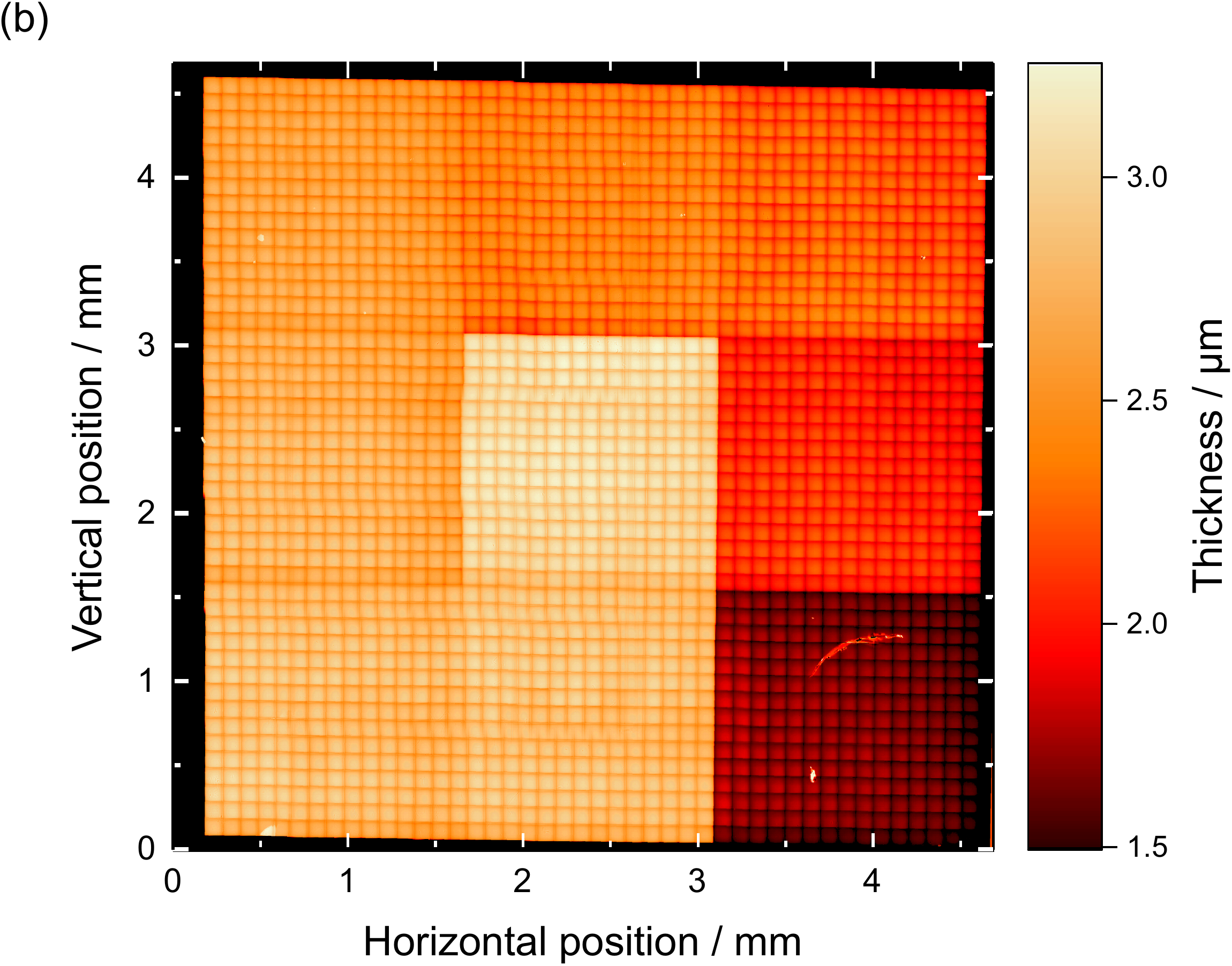}

    \caption{(a) 3D sketch of the static phase mask (z-dimension exaggerated). (b) Thickness of the phase mask measured with a confocal microscope. 
    }
    \label{fig4_mask}
\end{figure}

As a static demonstration object, a phase mask is fabricated on a silicon wafer (opaque to the near-infrared signal radiation) by three-dimensional laser lithography \cite{Hohmann2015} in a photoresist (IP-S from Nanoscribe \cite{Nanoscribe}). It consists of nine distinct cubes, each 1.5~mm square with the lowest and highest thickness of 1.56~µm and 2.87~µm, respectively. A 3D sketch of the phase mask is illustrated in Fig.~\ref{fig4_mask}(a).

Taking the index of refraction into account, the average thickness increments result in phase increments of about $0.3\pi$. For verification purposes, the thickness distribution of the phase mask was measured with a confocal microscope \cite{Nwaneshiudu2012} as shown in Fig.~\ref{fig4_mask}(b). Due to the employed fabrication technique, each of the nine squares consists of individual cubes with a footprint of 100~µm square, leading to the stitching edges observable in the thickness distribution. 

\begin{figure}[h]
    \centering
    \includegraphics[width=0.8\linewidth]{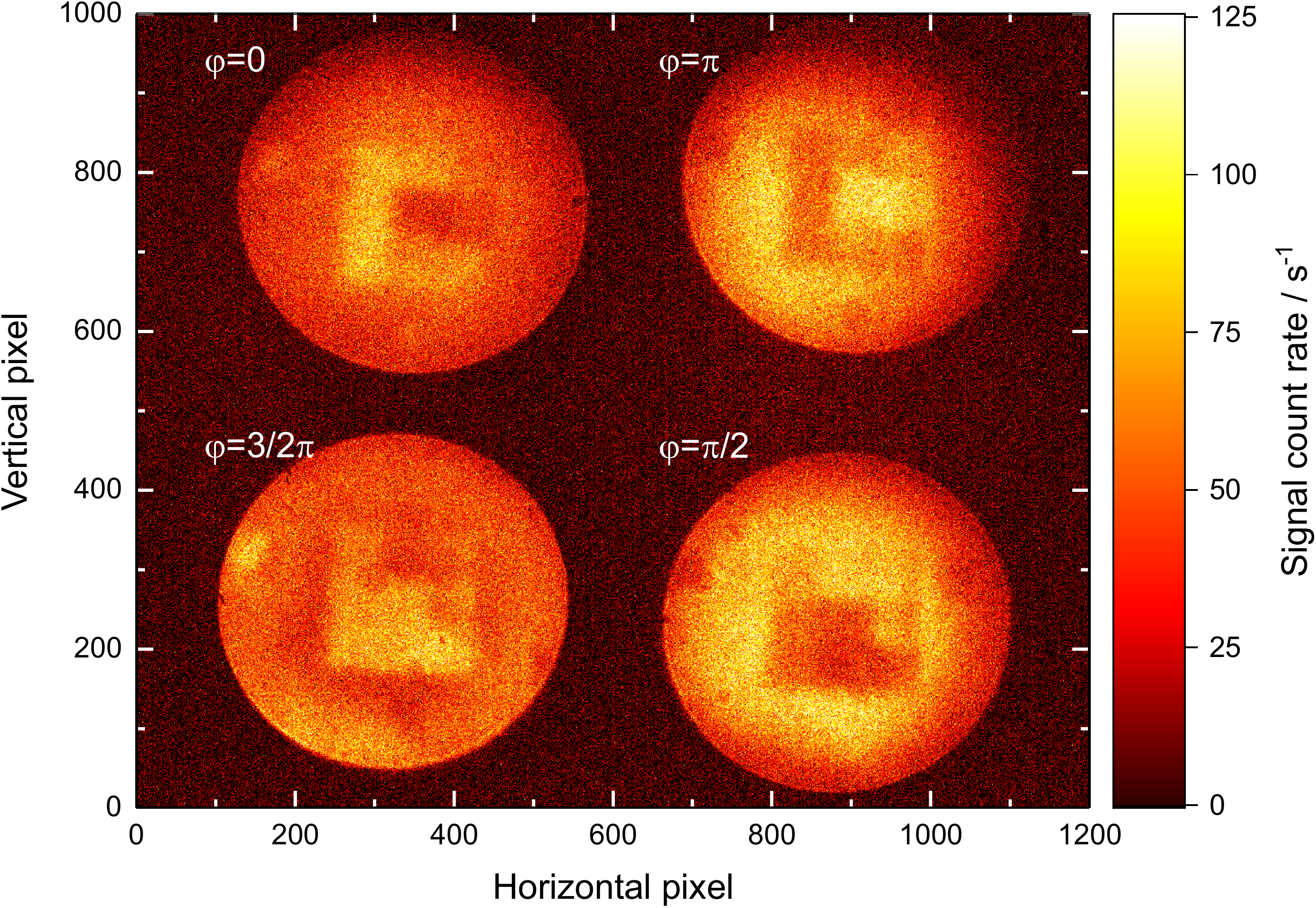}
    \caption{Signal spots with the phase mask placed in the idler beam path. Depending on the thickness of the mask the pixel-wise phase and therefore signal count rate differ.}
    \label{fig5_mask_quad}
\end{figure}

\begin{figure}[h]
    \centering
   
    \includegraphics[width=0.495\linewidth]{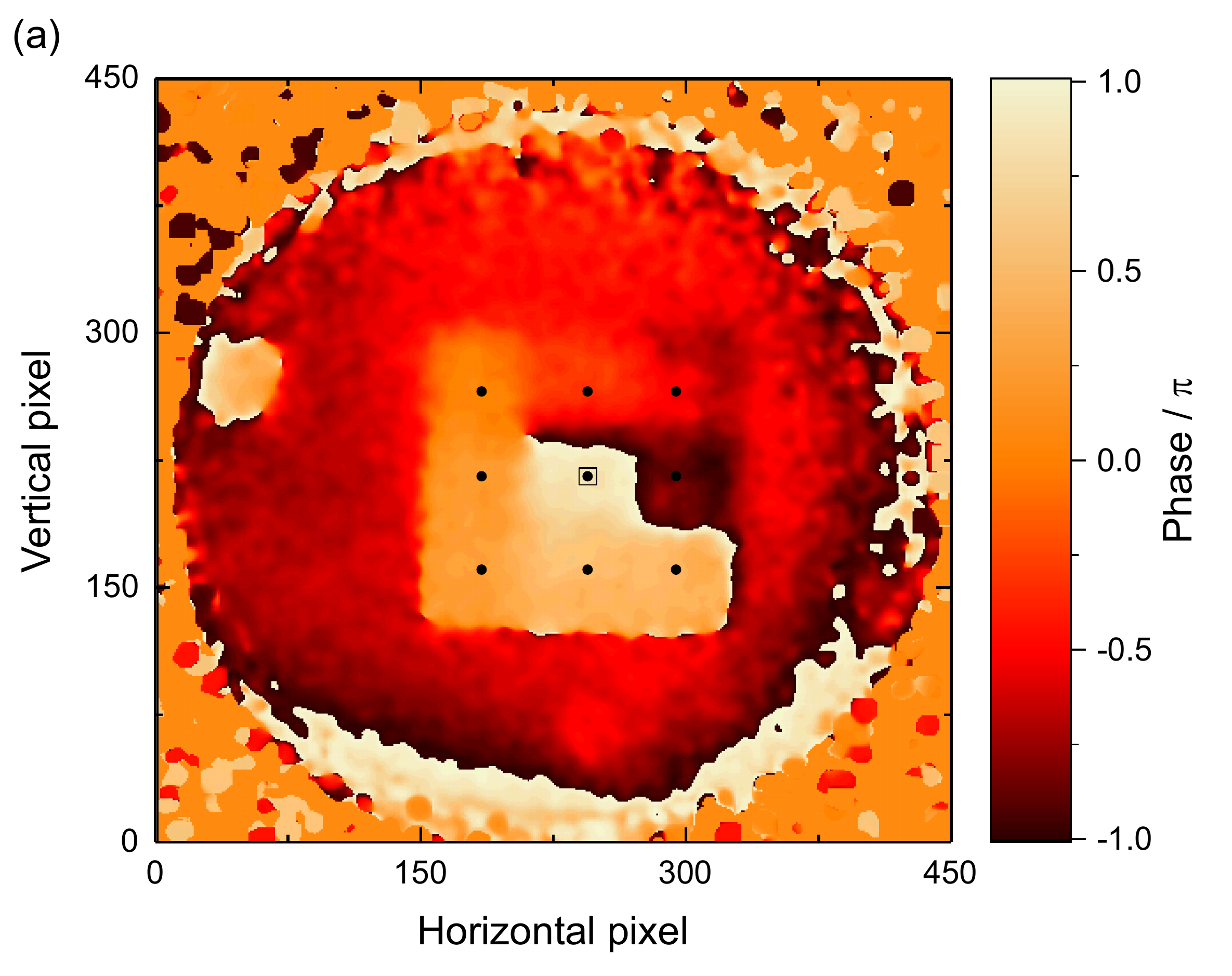}
    \includegraphics[width=0.495\linewidth]{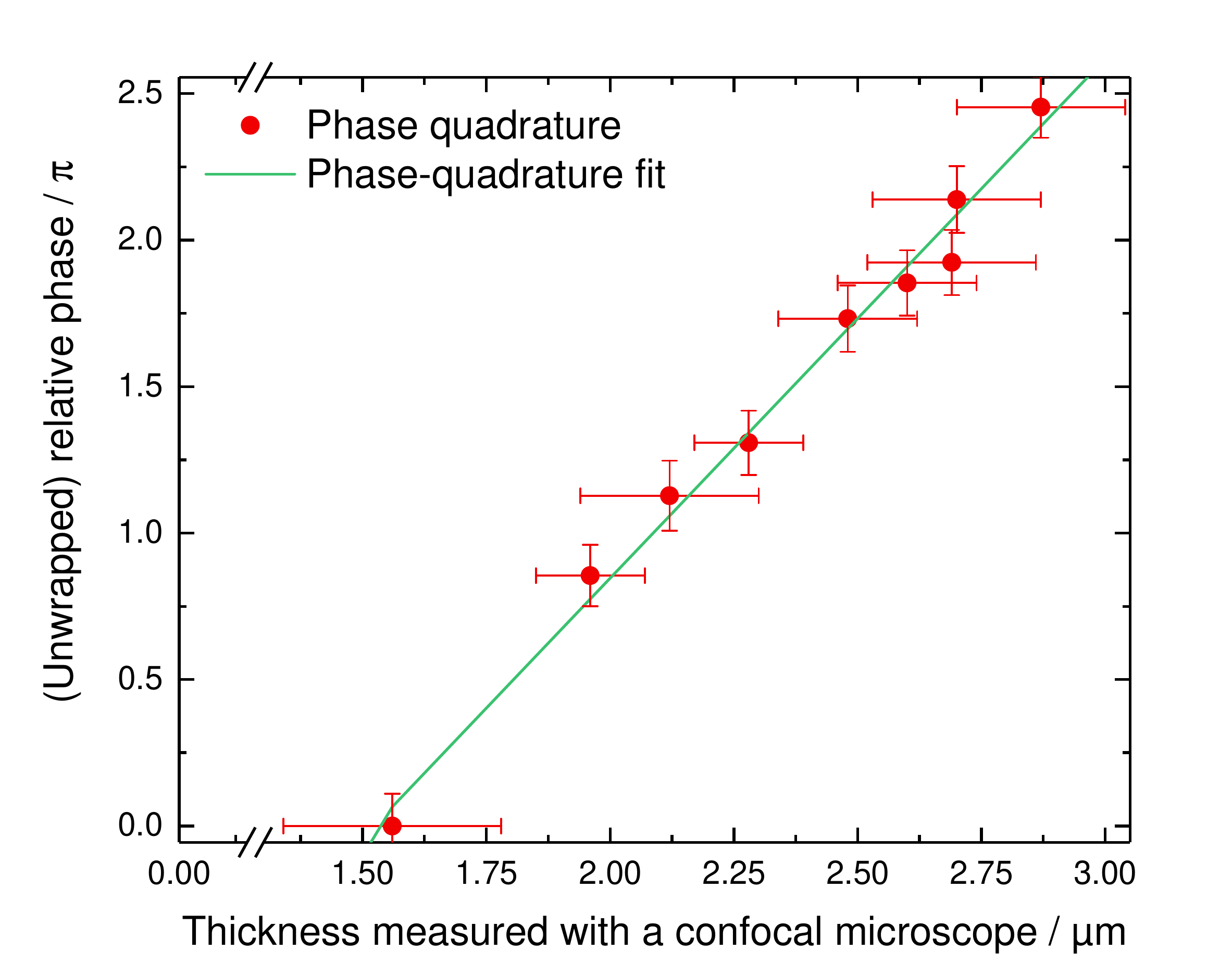}
    \caption{(a) Pixel-wise phase calculated by the evaluation of a single-shot phase-quadrature measurement. The dots indicate the positions within each of the phase-mask regions for which the phase was analyzed. (b) Relative phase of the different regions with different thicknesses of the phase mask with respect to the thinnest step.  The error bars refer to the standard deviation of the thickness measured with the confocal microscope and the standard deviation of the relative phases determined by the phase-quadrature recordings, respectively.}
    \label{fig6_mask_result}
\end{figure}

Figure~\ref{fig5_mask_quad} shows the raw camera image of the four signal spots with the phase mask introduced to the idler beam. At the left edge of each spot an additional small cube is visible used to verify the orientation of the phase mask. Slight tilts and mismatches in size can be corrected numerically, which was done in advance with a resolution target not shown here. Further, a spatial Gauss filter decreases noise. After this, the single image can be used to calculate the pixel-wise phase according to eq.~\ref{eq:Phase}, which is shown in Fig.~\ref{fig6_mask_result}(a). 

The helical-like structure of the phase mask is directly observable taking the phase wraps into account. Evaluating the unwrapped phase at the dots inserted in the figure, the correlation between our proposed method and the thickness measured with the confocal microscope is shown in Fig.~\ref{fig6_mask_result}(b). The coefficient of determination $r^{2}$ is $0.98$ and proves the usefulness of our proposed method.

\subsection{Dynamic sample measurements}

To demonstrate the advantage of our proposed method, time-varying measurements on dynamic samples are presented in the following, which would challenge the common measurement principle based on varying optical delays as discussed above. 

\begin{figure}[h]
    \centering
    \includegraphics[width=0.495\linewidth]{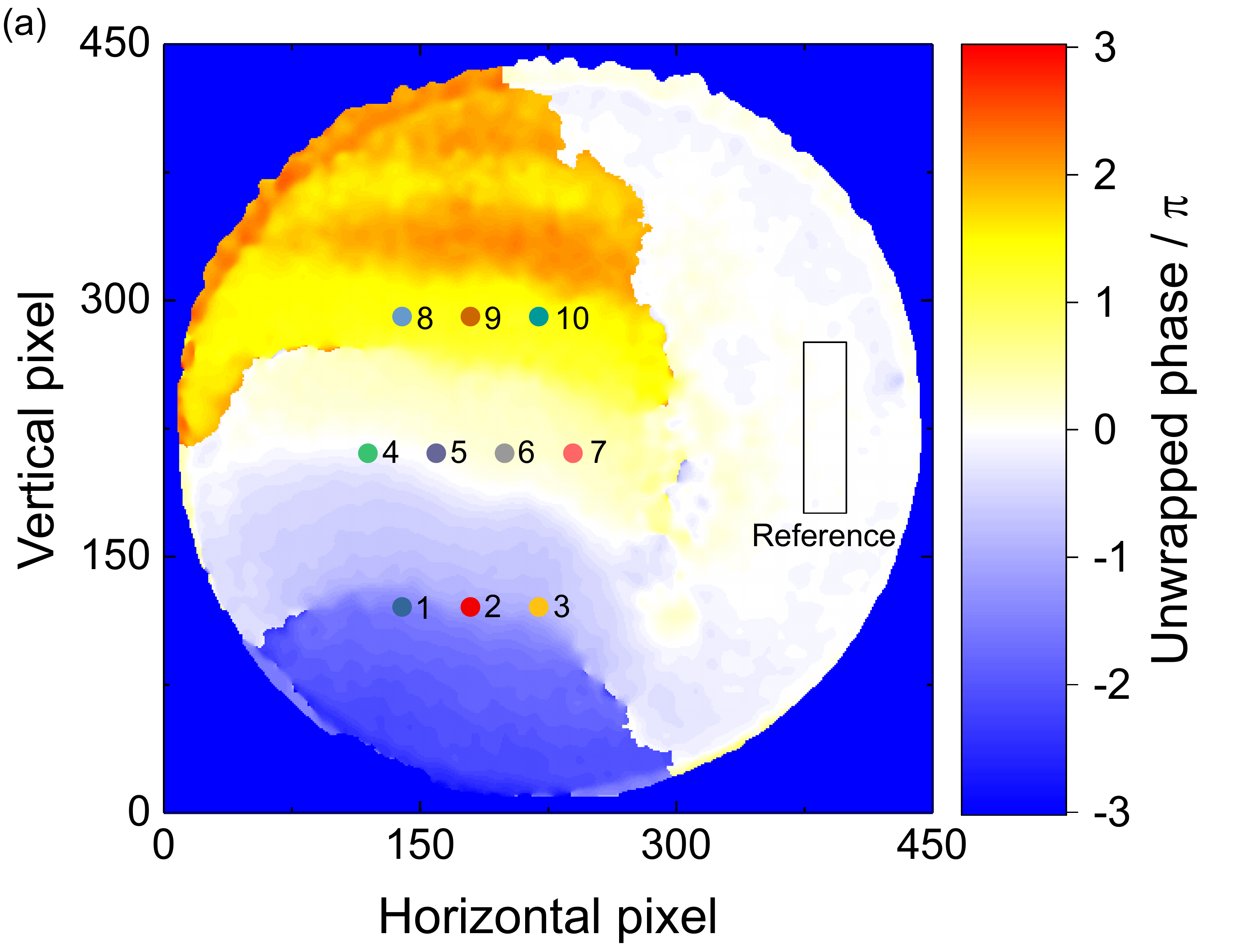}
    \includegraphics[width=0.495\linewidth]{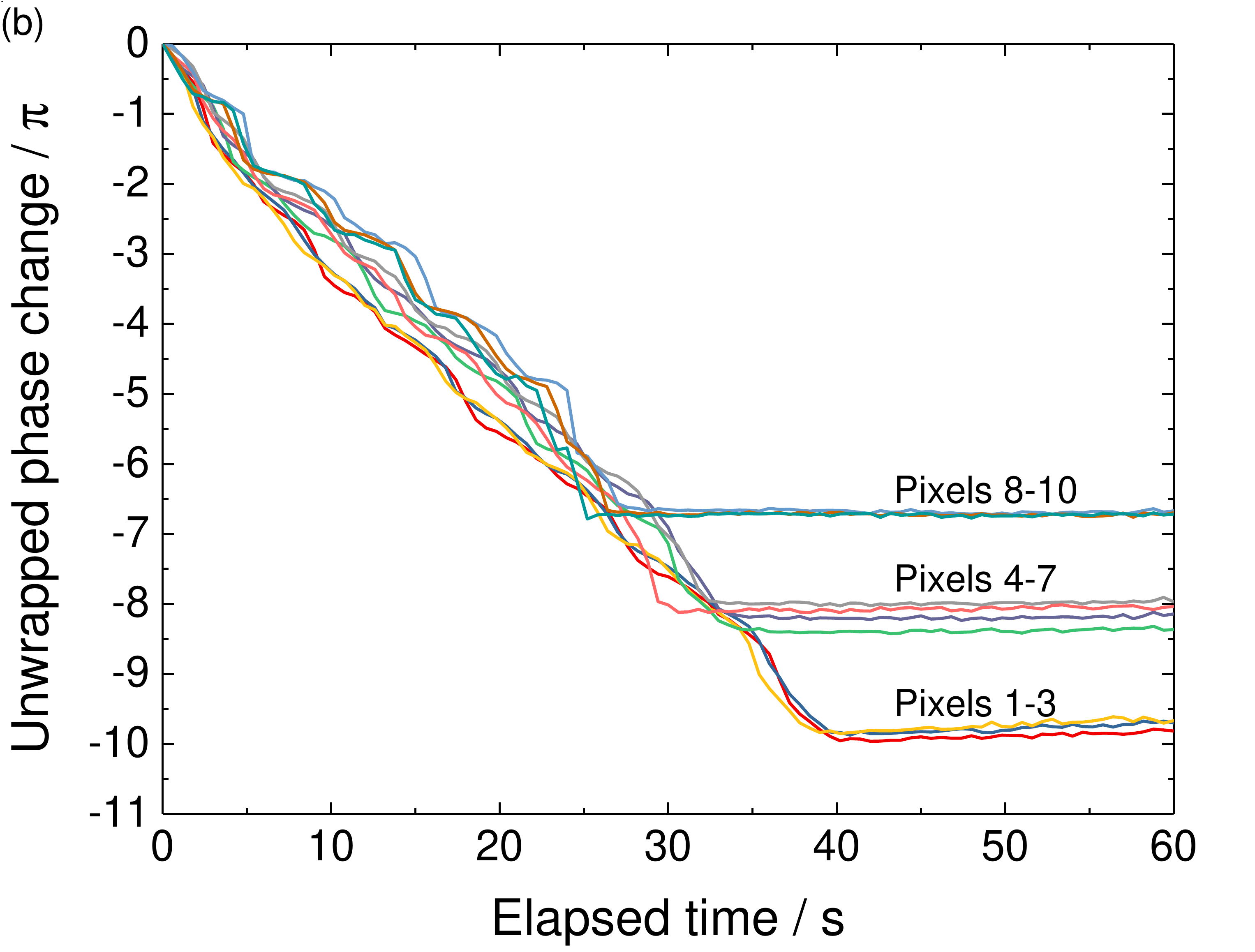}
    \caption{(a) Unwrapped phase after 15 seconds of the measurement time elapsed. The mean phase within the reference region is set to 0. (b) Unwrapped changing phase of a set of signal-spot areas (labeled pixels in (a)) in the course of elapsed time. }
    \label{fig7_isoprop}
\end{figure}

As a first dynamic example, the drying of a film of isopropanol on a glass substrate is observed by introducing it into the idler beam path and measuring the phase of the signal beam. During its evaporation, successive images are acquired and processed. In addition to the evaluation steps described above, the phase is unwrapped \cite{Herraez2002} to analyze the variance in the decreasing thickness of the film. An area of the FOV without sample is used to reference the phase, eliminating drifts of the interferometer itself. Fig.~\ref{fig7_isoprop}(a) shows the pixel-wise calculated and unwrapped phase change after a period of 15 seconds drying. 

The vertical phase distribution can be explained by gravity, which causes more isopropanol to accumulate at the bottom of the sample. Figure~\ref{fig7_isoprop}(b) shows the phase change over time for the signal point regions labeled in Fig.~\ref{fig7_isoprop}(a). A linear phase change is observed until the local isopropanol film is evaporated. The evaluated pixels show a grouping in their overall accumulated phase change according to their vertical position, underlining the distribution due to gravity. With the total phase change during the measurement of up to about $10\pi$ and the refractive index of isopropanol being 1.366 at the idler wavelength of about 1550\,nm \cite{Saunders2016}, we estimate the thickness of the isopropanol film of about 21~µm at the start of our measurement. 

\begin{figure}[h]
    \centering
    \includegraphics[width=0.495\linewidth]{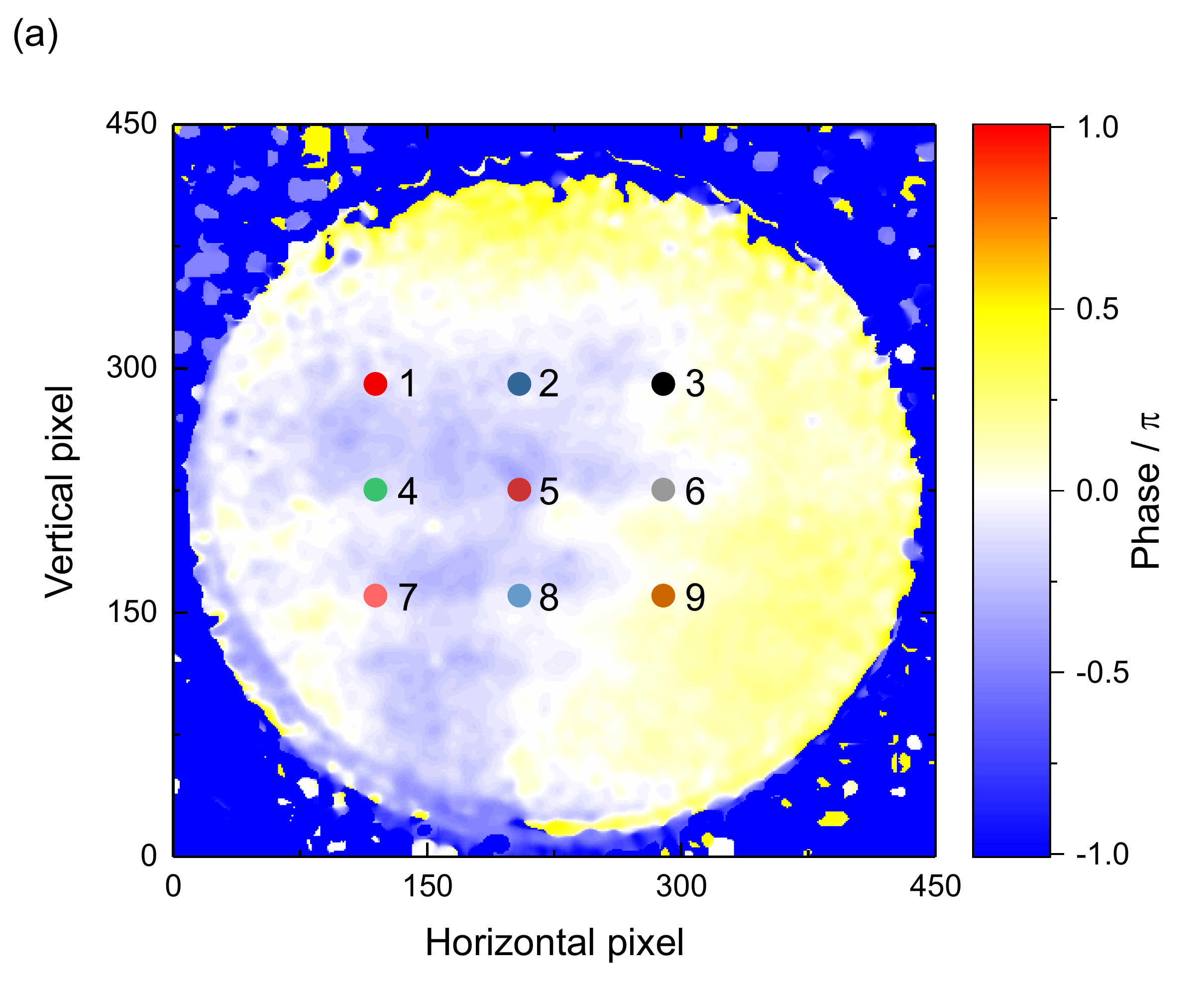}
    \hfill
    \includegraphics[width=0.495\linewidth]{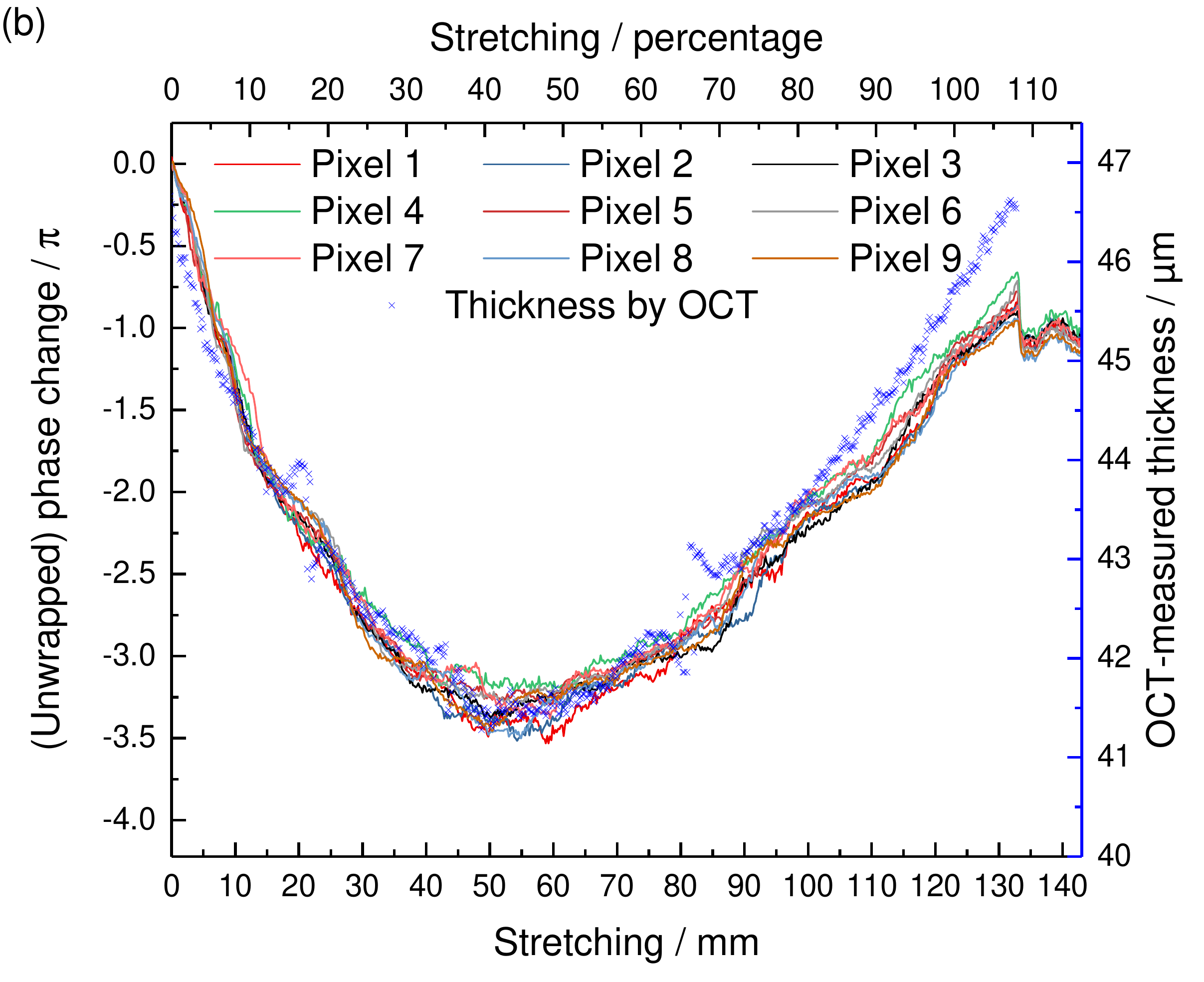}
    \caption{(a) Phase with the pixels for which the phase is determined in (b) after the tape is elongated 10~mm. (b) Changing phase over the course of time as the 123-mm-long scotch tape is stretched in comparison to a thickness measurement performed by a standard OCT system.}
    \label{fig8_tape}
\end{figure}

As another demonstration of dynamic measurement tasks, a 123-mm-long adhesive tape (tesafilm\textsuperscript{\textregistered} transparent) was placed in the idler beam path and stretched at a rate of 22 mm per minute. According to the manufacturer Beiersdorf, the tape consists of polypropylene with an adhesive layer sprayed on one side of it \cite{Beiersdorf}. To measure the strain-dependent change in optical thickness, many successive camera images are acquired during the stretching process. Figure~\ref{fig8_tape}(a) exemplarily shows the measured phase after stretching the tape by 10\,mm.

To verify the measurement results, the optical thickness is simultaneously recorded (next to the idler beam, but not overlapping with it) with a commercially available optical coherence tomography (OCT) system working in the visible and near-infrared spectral range. In the course of the stretching, this OCT-measured thickness is compared with the changing phase of some FOV positions in Fig.~\ref{fig8_tape}(b). As can be seen, both measurements show very similar results. In both measurements, the phase decreases at the beginning before reaching a minimum at an elongation of about 40\,\% and increasing again. Just before the strip breaks, it almost returns to the phase it had in the unstretched state at the beginning. However, the small differences, which become more pronounced as the tape is further stretched, could be due to the different wavelength ranges in which the measurements take place, since a dispersive stretching behavior of polypropylene produces differences in the measured phase.

The stretching-dependent change in the refractive index of the tape is divided into different ranges. According to the literature \cite{Samuels1979} in a rotation of the crystal lattice structure, followed by void formation. Finally, further stretching leads to deformation of the plastic itself. It can also be observed that the strip coils a little and its width decreases with increasing elongation. This could be the reason why the optical thickness and thus the phase of the tape increases again before the polypropylene tape finally breaks.

\subsection{Discussion}
All measurement results prove the applicability of our phase-quadrature measurement using single images without the need of varying optical delays. Obviously, the used illumination time of 500~ms is not state-of-the-art, but is only indebted by the limited performance of the used components (e.g. limited pump power of the employed fiber-coupled laser). Nevertheless, the advantage of the single-shot phase-quadrature measurement compared to concepts with varying optical paths is evident. In general, the proposed method is transferable to better performing nonlinear interferometers \cite{Gilaberte2021} and could directly enhance their dynamic measurement capability.

\section{Conclusions}
We have presented a technique for recording a complete quadrature of a nonlinear interference in a single image acquisition. This is achieved by applying the phase-quadrature detection principle using a combination of polarization optics and allows for a direct measurement of the phase and visibility of the nonlinear interference. In a proof-of-concept investigation, this measurement technique was used to measure the characteristics of a static phase mask placed in the idler beam path with a single image acquisition. To demonstrate the dynamic measurement capability, the drying process of a film of isopropanol and its effect on the phase of the signal interference as well as the change in refractive index of a stretched adhesive tape was measured. Classical measurement methods accompanying most of our experiments successfully verified our results. 

The presented method can be applied to nonlinear interferometers where the idler has a much longer wavelength and is in the fingerprint range in the mid-infrared \cite{Mukai2022, Kviatkovsky2022, Paterova2022} or even terahertz \cite{Kutas2020, Kutas2021} spectral range, and thus accelerating these interferometers as well. The same is true for the adaption to Michelson-like nonlinear interferometers, which are becoming increasingly popular \cite{Kutas2022}. Thus, this approach can enable or enhance the measurements of dynamic processes and improve the future competitiveness of nonlinear interferometers \cite{Gilaberte2021, Kviatkovsky2020, Lindner2020, Paterova2020}.

\begin{backmatter}
\bmsection{Funding}
Fraunhofer-Gesellschaft (Lighthouse project QUILT).

\bmsection{Acknowledgments}
B.H., J.H., M.K., G.v.F. and D.M. acknowledge the Fraunhofer-Gesellschaft for funding during the Lighthouse project QUILT (Quantum Methods for Advanced Imaging Solutions). We thank Felix Riexinger for helpful discussions.

\bmsection{Disclosures}
The authors declare no conflicts of interest.

\bmsection{Data availability} Data underlying the results presented in this paper are not publicly available at this time but may be obtained from the authors upon reasonable request.

\end{backmatter}

\bibliography{references}

\end{document}